# Adaptive Fusion Graph Network for 3D Strain Field Prediction in Solid Rocket Motor Grains

HUANG Jia-da[1,2] , MA Hao[3, *], SHEN Zhi-bin[1,2, *], QIAO Yi-zhou[1,2], LI Hai-yang[1,2]

[1] College of Aerospace Science and Engineering, National University of Defense Technology, Changsha, 410073, China

[2] State Key Laboratory of High-Efficiency Reusable Aerospace Transportation Technology

[3] School of Aerospace Engineering, Zhengzhou University of Aeronautics, 450046 Zhengzhou, China

[*]Corresponding author

**Abstract:**

Local high strain in solid rocket motor grains is a primary cause of structural failure. However, traditional numerical simulations are computationally expensive, and existing surrogate models cannot explicitly establish geometric models and accurately capture high-strain regions. Therefore, this paper proposes an adaptive graph network, GrainGNet, which employs an adaptive pooling dynamic node selection mechanism to effectively preserve the key mechanical features of structurally critical regions, while concurrently utilising feature fusion to transmit deep features and enhance the model's representational capacity. In the joint prediction task involving four sequential conditions--curing and cooling, storage, overloading, and ignition--GrainGNet reduces the mean squared error by 62.8% compared to the baseline graph U-Net model, with only a 5.2% increase in parameter count and an approximately sevenfold improvement in training efficiency. Furthermore, in the high-strain regions of debonding seams, the prediction error is further reduced by 33% compared to the second-best method, offering a computationally efficient and high-fidelity approach to evaluate motor structural safety.

## 1. Introduction

Solid Rocket Motors (SRMs), the core propulsion systems of solid-fueled missiles, play a critical role in ensuring missile safety and reliability through their structural integrity[1-3]. The grains, the core component of SRMs, display pronounced nonlinear mechanical behavior under realistic operating conditions due to their viscoelastic material properties and complex geometric structure[4,5]. Local high-strain zones often develop in particular regions (such as the roots of star-shaped grain grooves), potentially leading to structural failure[6-8]. Therefore, the rapid and accurate prediction of strain distribution in these critical regions is essential for grain structural optimisation and failure prevention[1,9].

However, in the initial stages of engine design, traditional numerical simulations necessitate solving highly nonlinear governing equations for the complex geometry of the grain, resulting in single simulations that can take hours or even days[10-14]. While experimentation (such as fiber optic sensor measurements[15,16]) can capture actual physical responses, they are constrained by limited data dimensionality and high costs. This dual bottleneck of simulation and experiment severely restricts engineering efficiency, particularly in scenarios that require rapid iteration, such as design optimisation, parameter inversion, or real-time health monitoring.

Although traditional surrogate models (such as Gaussian process regression[17], RBF models[18,19]) can accelerate the prediction of the grain's mechanical response, they fundamentally map design parameters to responses and cannot explicitly model the grain's geometric structure. In design optimisation, these "geometry-agnostic" models can only predict discrete data points and are limited to reconstruct full-field strain distributions.

With the advancement of artificial intelligence techniques, deep neural networks have demonstrated strong potential in constructing high-dimensional, nonlinear mapping relationships, offering new approaches to overcome the bottlenecks mentioned above[20-22]. Among these methods, the Multi-Layer Perceptron (MLP), as a basic fully-connected architecture, is advantageous in characterising the global nonlinear mapping of design parameters and complex responses[23], but its inherent lack of topology-awareness makes it challenging to process spatial structural information. On the other hand, although Convolutional Neural Networks (CNNs) can extract spatial features, they heavily rely on structured mesh inputs[24-27] and require discretising the irregular geometry of the propellant into a regular pixel mesh, which not only introduces geometric errors but also reduces strain resolution in critical regions[28].

Graph Neural Networks (GNNs) are naturally suited to finite element unstructured grids[29] due to their ability to process non-Euclidean data(such as social networks[30] and chemical molecules[31] ) along with arbitrary topological connections[32], making them key to overcoming the bottlenecks mentioned above: GNNs directly map finite element meshes of grain to graphs [33], fully preserving geometric features and avoiding the resolution loss from conversion to regular tensors[28,34]. This technology has been successfully applied in computational mechanics, including composite damage prediction, metal fatigue crack propagation, and aerodynamic optimization[35-40]. Specifically, Maurizi et al. [41] proposed an improved graph network method to map material microstructure to physical response; Wang et al. [42] developed a global-information-guided graph network to achieve nonlinear mapping from feature parameters to airfoil flow fields and hub pulley pressure fields; Li et al.[43] built an aerodynamic strength prediction graph network to predict physical fields of gas turbines under different boundary conditions. However, the above GNN-based mechanics response studies still face two major challenges: feature transmission degradation between layers in deep networks leads to signal attenuation in high-strain regions (such as star tip transition zones), and uniform neighbor aggregation strategies lack sufficient sensitivity to local critical areas, which will be discussed in detail later.

Therefore, this study proposes a Grain Graph Network (GrainGNet) framework for rapid prediction of strain fields in solid rocket motors. Our framework introduces two key innovations:

1) Adaptive sampling module: dynamically focuses on high-strain regions to enhance local sensitivity.

2) Adaptive feature fusion module: employs cross-layer skip connections to suppress feature degradation in deep networks.

To our knowledge, this is the first use of graph neural networks for strain field prediction in SRMs reported in the literature. The paper is structured as follows: Section 2 establishes the parameterized model of SRM grain and graph data construction methodology; Section 3 details the GrainGNet architecture; Section 4 presents systematic validation with quantitative metrics and strain field predictions; Finally, Section 5 concludes the study (overview in Fig. 1).

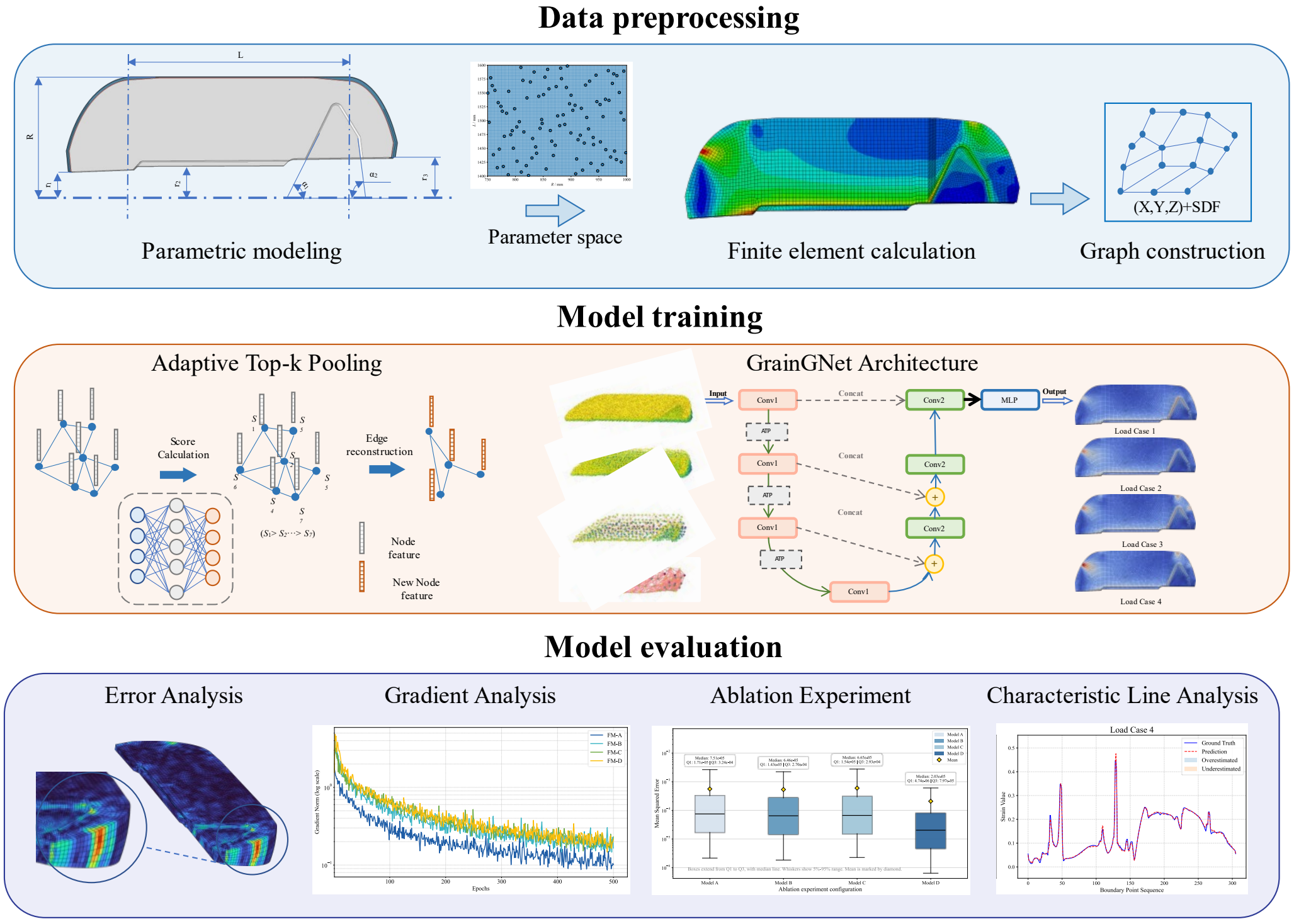

Fig. 1 Research flowchart.

## 2. Dataset construction

### 2.1 Parametric modelling

This study uses the rear-wing star-hole grain as a case study. Considering the cyclic symmetry of both the geometric model and the external loads (such as temperature and internal pressure) it experiences, a 1/12 circumferential geometric model of the motor is selected as the analysis object. To simplify the modeling process, components such as the nozzle and igniter, as well as process details, were omitted. Finally, based on a parametric modeling approach, a finite element model of the motor was constructed with the shell, insulation layer, cladding layer, and grain as the main components, as shown in Fig. 2.

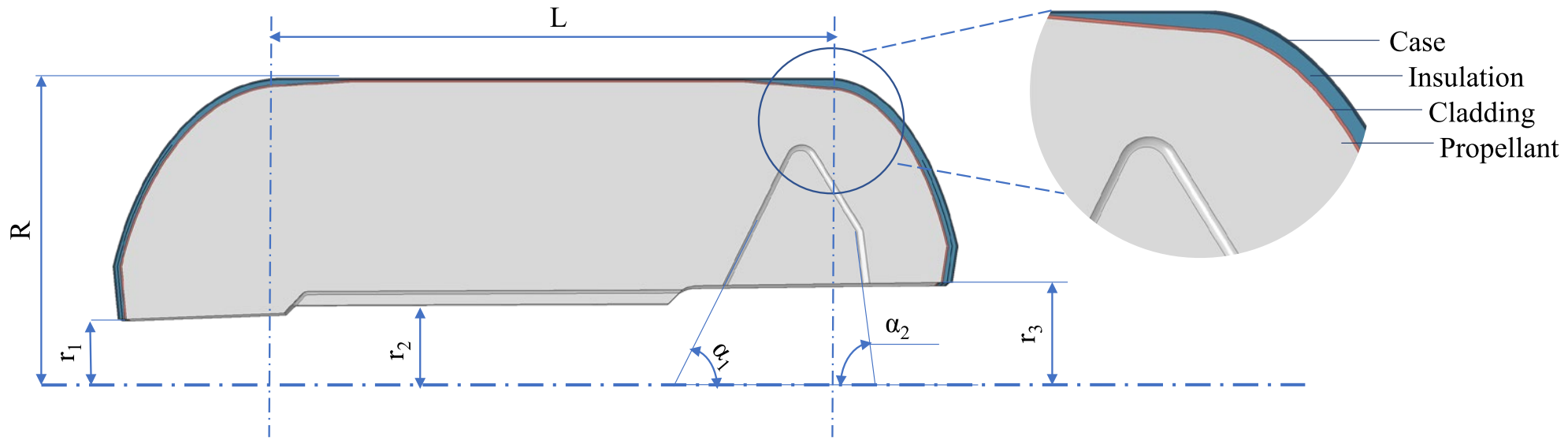

Fig. 2 1/12 parameterised SRM model

Regarding the application of boundary conditions, due to the structural symmetry of the motor, the circumferential displacements of the two symmetric faces constraining the combustion chamber are restricted, and displacement constraints are applied to the outer surface of the shell. Additionally, to reduce stress concentration, artificial debonding seams are introduced at the head and tail.

The material parameters for the various parts of the motor are shown in Table 1, where they represent the modulus $E$, Poisson's ratio $\nu$, thermal expansion coefficient $\alpha$, and density $\rho$, respectively. Given the pronounced viscoelastic characteristics exhibited by the propellant, its mechanical behavior shows typical time and temperature dependence, and traditional elastic or elastoplastic constitutive models cannot accurately capture these time-varying properties. Therefore, this study is based on the generalised Maxwell viscoelastic model framework. The shear relaxation modulus master curve is obtained through relaxation tests and fitted using a 3-term Prony series expansion. The shear relaxation modulus $G(t)$ for the cladding layer and the grain is:

$$G(t) = 1.969 - 0.6316(1 - e^{-t/5.506}) - 0.4405(1 - e^{-t/55.06}) - 0.4456(1 - e^{-t/550.6}) \tag{1}$$

where $t$ is the relaxation time. At the same time, to characterise the effect of temperature on material performance, the Williams-Landel-Ferry (WLF) equation is used to describe the time–temperature shift factor $\alpha_T$ :

$$\lg \alpha_T = \frac{-14.19(T - 293.15)}{173.46 + (T - 293.15)} \tag{2}$$

where $T$ is the thermodynamic temperature.

Table 1 Material parameters for SRM

| Parts | $E$/MPa | $\nu$ | $\alpha$/(1/K) | $\rho$/(kg/m$^3$) |
|---|---|---|---|---|
| Shell | $1.86\times10^5$ | 0.3 | $1.1\times10^{-5}$ | 7900 |
| Insulation | 22.0 | 0.4985 | $2.2\times10^{-5}$ | 2100 |
| Cladding/Propellant | $E(t)$ | 0.498 | $8.6\times10^{-5}$ | 1151 |

To meet the needs of strain-field prediction throughout the motor's entire life cycle, the motor is considered to undergo several conditions: curing cooling, vertical storage, overload, and ignition boost. The specifics are as follows:

1） Curing and cooling condition: From the zero-stress temperature 58 °C, the motor is cooled over 24 h to 20 °C;

2） Vertical storage condition: A gravitational acceleration is applied axially to the entire model, sustained for half a year;

3） Ignition pressurisation condition: An internal pressurisation load of 5 MPa is applied to the inner surface of the combustion chamber, lasting 0.3 s.

4） Overload condition: An acceleration load directed from the front end toward the rear end is applied to the model, increasing within 0.8 s to 10 g;

## 2.2 Dataset generation

This study focuses on the grain. Based on the principle of independence, as shown in Table 1, seven key geometric parameters are selected as variables—such as the radius of the head cross-section ( $r_1$ ) and the starting angle of the rear wing( $\alpha_1$ )—while other structural parameters are kept fixed, to analyse the effects of these key geometric parameters on the strain field distribution of the grain.

To improve the sampling efficiency in the multi-dimensional parameter space, this study employed optimal Latin hypercube sampling for experimental design. This method first evenly partitions each parameter's range into 200 intervals and randomly selects one value from each interval. Through multiple iterations, the sampling scheme that maximises the minimum distance between any two points is selected, thereby balancing hierarchical coverage and the maximum interval between points. As illustrated in Fig. 3, with the two-dimensional projection example of parameters *R* and *L*, the sampled points exhibit uniform coverage across the entire space. Based on these parameter combinations, finite element simulations were used to obtain the corresponding strain field data, constructing a high-quality training dataset capable of accurately characterising the complex nonlinear mapping between geometric parameters and physical fields, and providing reliable training samples for deep learning models.

Table 2 Range of Variables

| Variable | $R$ / mm | $L$ / mm | $r_1$ / mm | $r_2$ / mm | $r_3$ / mm | $\alpha_1$ / ° | $\alpha_2$ / ° |
|---|---|---|---|---|---|---|---|
| Sampling Space | [750,1000] | [1400,1600] | [120,140] | [160,200] | [220,260] | [50,70] | [70,90] |

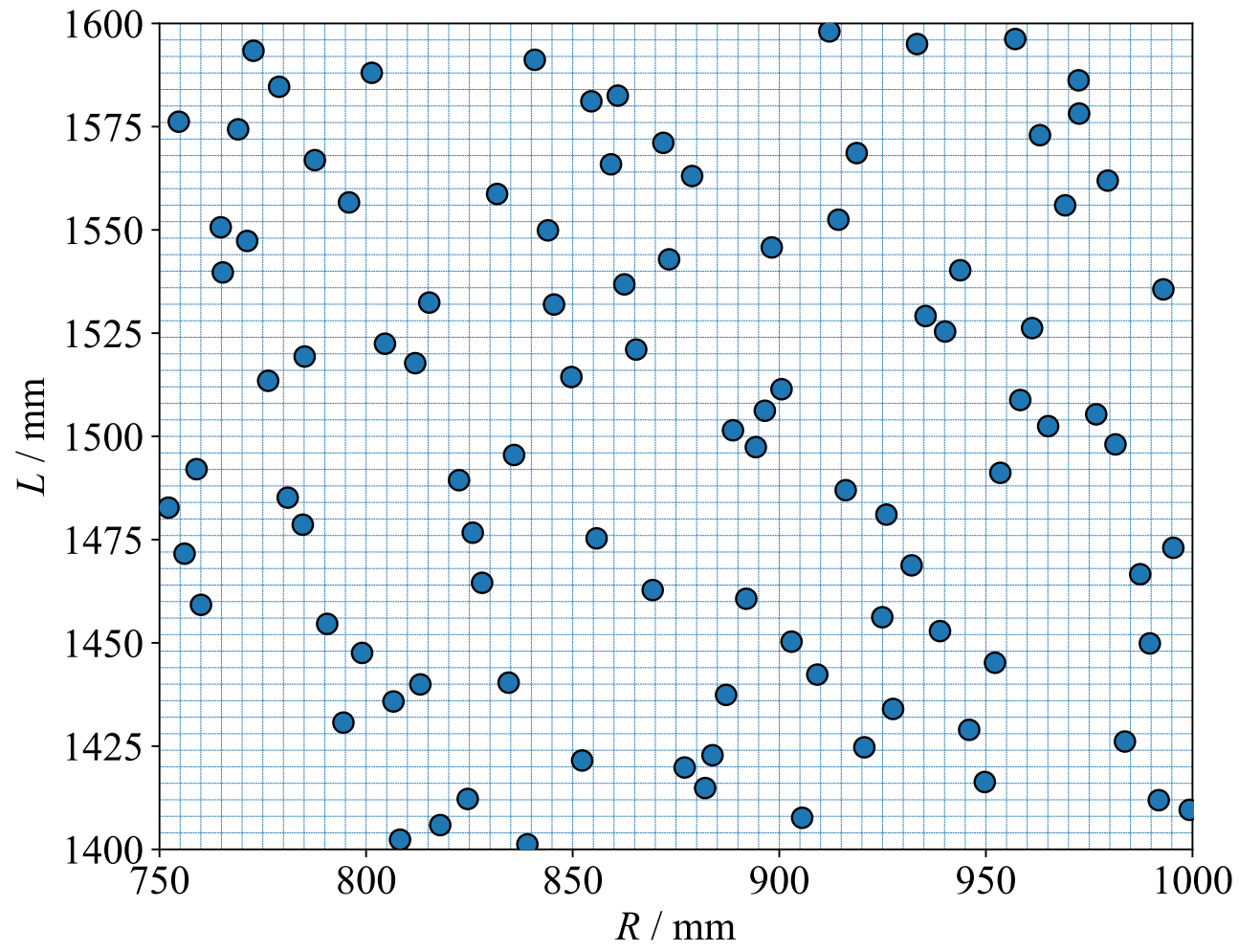

Fig. 3 Optimal Latin hypercube sampling (taking the two-dimensional projection of parameters *R* and *L* as an example).

The parametric modelling is based on a Python script that automates the process through the ABAQUS API interface. The script automatically generates a 3D finite element model containing the shell, insulation, cladding, and grain (Fig. 2) based on the seven geometrical parameters entered (Table 2). C3D8RH elements were used for meshing, and the strain field was solved through implicit dynamic analysis. After the batch calculation, the 3D coordinates of each node, the corresponding Mises strain values for the four working conditions, and the element connection relationships were extracted.

## 2.3 Graph input and preprocessing

Based on the 3D coordinates $(x, y, z)$ as the input features for the graph neural network, this paper further introduces the Signed Distance Function (SDF) to enhance the modelling capability of geometric information. For the crack-free structure, the region of higher strain often appears near the free boundary. Based on this physical phenomenon, the signed distance from a point to the structural boundary is used as an additional feature that helps the graph network capture the strain trends near the boundary. The signed distance function is defined as follows:

$$\phi(x) = \begin{cases} \text{dist}(x, \partial\Omega), & \text{if } x \in \Omega \\ -\text{dist}(x, \partial\Omega), & \text{if } x \notin \Omega \end{cases} \tag{3}$$

where $\Omega \subset \mathbb{R}^3$ denotes the structural domain, $\partial\Omega$ denotes its boundary, and $\text{dist}(x, \partial\Omega)$ denotes the Euclidean distance from the point $x$ to the boundary.

The input graph $G^{(0)} = (V^{(0)}, E^{(0)})$ consists of a set of nodes $V^{(0)}$ and a set of edges $E^{(0)}$. The initial features $x_i^{(0)} \in \mathbb{R}^d$ of each node $i$ include the 3D coordinates, the SDF value of the signed distance function, and the Mises strain labels. The edges $E^{(0)}$ are generated using the Delaunay triangulation method to reflect the spatial connectivity of the nodes.

Based on the processed graph data, a hierarchical random sampling strategy is used to divide the dataset into a training set (70%), a validation set (15%), and a test set (15%), constructing the hierarchical data structure containing the original graph and cluster mapping relations.

## 3. Network architecture

In this work, as shown in Fig. 4, a deep neural network architecture named GrainGNet is proposed. This architecture adopts the U-Net structure as its framework, utilises hierarchical graph convolution operations to extract and reconstruct geometric features, and innovatively introduces the Adaptive Top-k pooling (ATP) module and the Adaptive Feature Fusion(AFF) module, significantly enhancing the representation capability of complex grain structures. The network input consists of node coordinates and SDF information from the unstructured grid of the grains, and the output provides full-field performance prediction results under multiple load cases. The core framework is illustrated in Fig. 4.

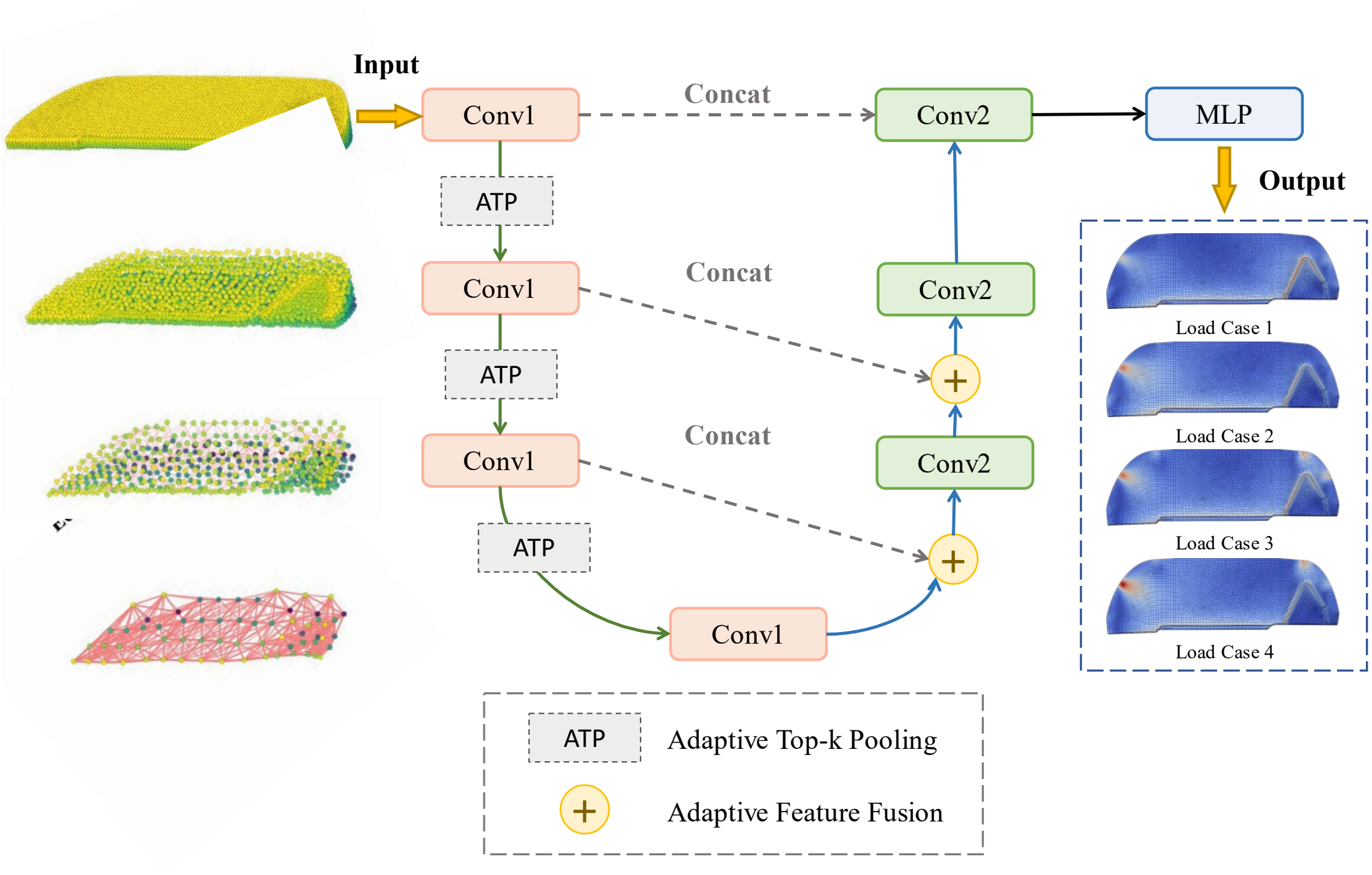

Fig. 4 GrainGNet

### 3.1 Adaptive Top-k Pooling (ATP)

In the downsampling stage, due to the non-uniformity and complex geometry of grain meshes, conventional fixed-coarsening downsampling struggles to satisfy the accuracy requirements in critical regions. Although inspired by Cangea[44] to adopt Top-k pooling, that approach relies solely on simple dot-product learning of linear features, and the occurrence of index discontinuities during graph reconstruction further increases the complexity of subsequent operations.

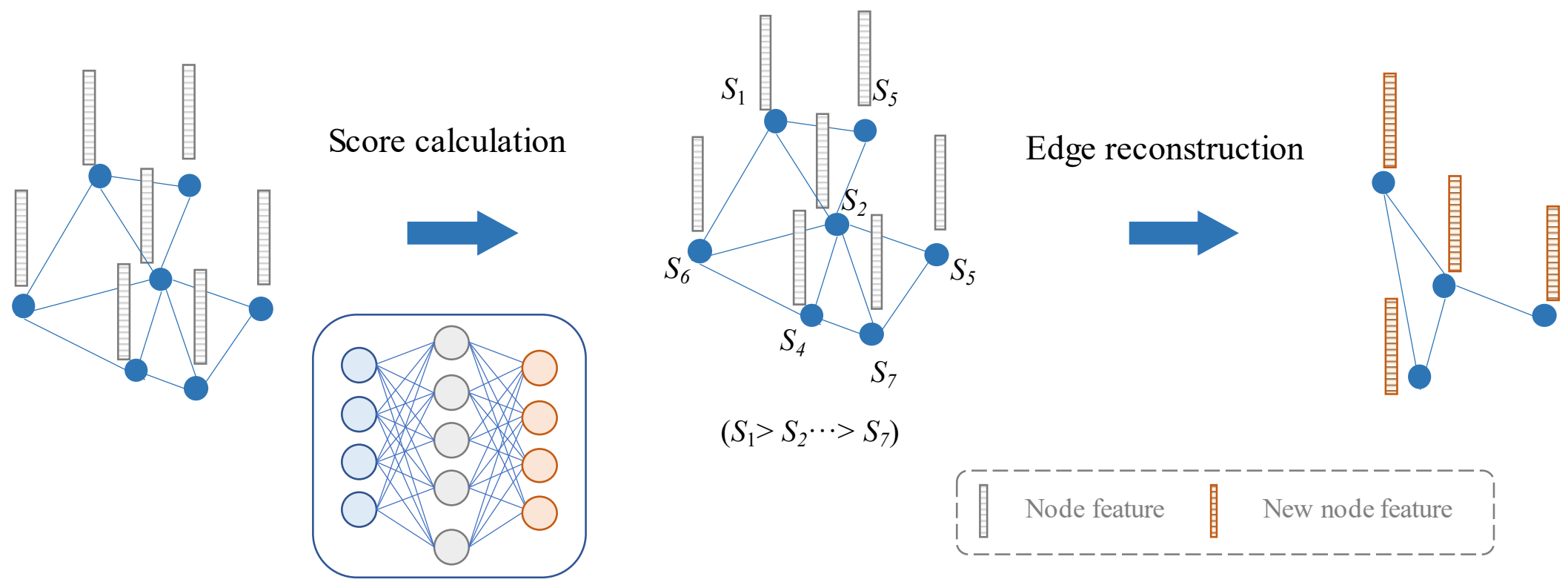

Fig. 5 The architecture of the proposed adaptive Top-k pooling.

For this reason, this study proposes the ATP method. The core of this approach lies in emulating the expertise of motors in identifying critical regions by introducing a lightweight "value assessment module" implemented as an MLP. This module replaces the conventional dot product, enabling intelligent selection of high-value nodes (e.g., high-strain regions) and precise data compression, akin to highlighting key points. The overall methodology is depicted in Fig. 5. The ATP method is realised through a lightweight MLP module comprising two fully connected layers. The first layer maps node features from a d-dimensional space to a latent space, followed by the application of the ReLU activation function. The second layer generates scalar importance scores. Specifically, for a given node $v_i$ , its importance score $s_i$ is calculated as follows:

$$s_i = \mathbf{W}_2^{\mathrm{T}} \times \mathrm{ReLU}\left(\mathbf{W}_1^{\mathrm{T}} \times x_i + b_1\right) + b_2 \tag{4}$$

where $x_i \in \mathbb{R}^d$ is the node feature vector, $\mathbf{W}_1 \in \mathbb{R}^{d\times 128}$ and $\mathbf{W}_2 \in \mathbb{R}^{128\times 1}$ are the learnable weight matrices, $b_1 \in \mathbb{R}^{128}$ and $b_2 \in \mathbb{R}$ are the bias terms. During network training, this evaluation module is jointly optimised with the main model by end-to-end gradient backpropagation, so that the importance score $s_i$ dynamically reflects the node's contribution to the multi-condition prediction task.

In the inter-layer feature compression stage, Top-k pooling based on importance scores is used to achieve adaptive downsampling. Specifically, the top $k$ nodes with higher importance scores are retained, and the number of retained nodes is adjusted according to a preset sampling rate. In this way, the pooled node set $\mathcal{V}_{\text{pool}}$ will contain the nodes that have the most significant impact on the prediction results, expressed by the formula:

$$\mathcal{V}_{\text{pool}} = \underset{|\mathcal{V}'|=k}{\arg\max} \sum_{i\in\mathcal{V}'} s_i, \quad k = \lceil \rho N \rceil \tag{5}$$

where $N$ is the total number of nodes in the current stratum, $\rho \in (0,1)$ is the preset sampling rate, and $\lceil \cdot \rceil$ denotes upward rounding.

### 3.2 Adaptive Feature Fusion(AFF)

To effectively integrate the characteristics of global deformation patterns and local strain distributions, the AFF module is developed, building upon the work of Lin[45]. The core concept of this module is to intelligently combine global patterns captured in deep features with local details preserved in shallow features, thereby generating a comprehensive and precise prediction.

The AFF module employs a pyramid-like structure for feature fusion. Starting from the coarsest-grained layer, features are progressively fused layer by layer upwards until the specified fusion depth is reached (three layers in this model). Specifically, at each layer, the fused deep-layer features are upsampled to match the resolution of the preceding (finer-grained) layer. These are then concatenated with the features of the current layer and subsequently fused using an MLP. Upsampling is achieved through an index mapping that records the clustering relationships between fine-grained and coarse-grained nodes.

Let the feature layer list be $\mathcal{F} = \{\mathbf{F}^{(0)}, \mathbf{F}^{(1)}, \ldots, \mathbf{F}^{(L-1)}\}$ , where $\mathbf{F}^{(0)}$ the finest-grained layer (input layer), $\mathbf{F}^{(L-1)}$ the coarsest-grained layer, $\mathbf{F}^{(i)} \in \mathbb{R}^{N_i \times d}$ , $N_i$ is the number of nodes in the $i$-th layer, and $d$ is the feature dimension. The list of clustering mappings $\mathcal{C} = \{\mathbf{C}^{(0)}, \mathbf{C}^{(1)}, \ldots, \mathbf{C}^{(L-2)}\}$ , where $\mathbf{C}^{(i)} \in \mathbb{Z}^{N_i}$ the mapping is from one $i$ layer to another $i+1$ layer. $C_j^{(i)} = k$ Indicates that the node $j$ in layer $i$ is mapped to the node $k$ in layer $i+1$ . The fusion process starts from the coarsest layer $\mathbf{F}^{(L-1)}$ , and then fuses to finer layers until it reaches the fusion depth $D$ ($D = 3$ ). The fused features are used in the subsequent up-sampling and decoding process. For the specific implementation process, please refer to the following pseudocode:

---

**Algorithm 1:** Adaptive Feature Fusion

---

**Input:**
$\mathcal{F} = \{\mathbf{F}^{(0)}, \mathbf{F}^{(1)}, \ldots, \mathbf{F}^{(L-1)}\}$ : Feature matrices.
$\mathcal{C} = \{\mathbf{C}^{(0)}, \mathbf{C}^{(1)}, \ldots, \mathbf{C}^{(L-2)}\}$ : Cluster mappings.
D : Fusion depth
**Output:** $\mathbf{H}^{(L-D)}$ : Fused feature matrix

// Initialize with coarsest features
$\mathbf{H}^{(L-1)} \leftarrow \mathbf{F}^{(L-1)}$

// Pyramid fusion: bottom-up traversal
for $i$ from $(L-2)$ to $(L-D)$ do:
    // Upsampling features via cluster mapping
    $\mathbf{U}^{(i)} \leftarrow \mathbf{H}^{(i+1)}[\mathbf{C}^{(i)}]$ //$[\cdot]$ represents index operation

    // Resolution alignment
    if $\| \mathbf{U}^{(i)} \|_0 < N_i$ then
        //Projection: truncate or zero-pad
        $\mathbf{U}_{\text{align}}^{(i)} \leftarrow \begin{bmatrix} \mathbf{U}^{(i)} \\ \mathbf{0}_{(N_i - \| \mathbf{U}^{(i)} \|_0) \times d} \end{bmatrix}$
    else $\mathbf{U}_{\text{align}}^{(i)} \leftarrow \mathbf{U}^{(i)}[:N_i]$ // Truncation processing

---

// Feature integration

$\mathbf{Z}^{(i)} \leftarrow \left[\mathbf{U}_{\text{align}}^{(i)} \,\|\, \mathbf{F}^{(i)}\right]$

// Fusion function

$\mathbf{H}^{(i)} \leftarrow \mathrm{ReLU}\left(\mathrm{MLP}_{\theta}\left(\mathbf{Z}^{(i)}\right)\right)$ // $\mathrm{MLP}_{\theta}(\mathbf{X}) = \mathbf{W}_2 \cdot \mathrm{ReLU}(\mathbf{W}_1 \cdot \mathbf{X} + \mathbf{b}_1) + \mathbf{b}_2$ ,

// $\mathbf{W}_1 \in \mathbb{R}^{d\times 2d}$ $\mathbf{W}_2 \in \mathbb{R}^{d\times d}$

**return** $\mathbf{H}^{(L-D)} \in \mathbb{R}^{N_{L-D}\times d}$

## 4. Results and discussion

### 4.1 Experimental setup

The training process employed the Adam optimiser with a fixed learning rate of 0.001 over 500 epochs, utilising an NVIDIA GeForce RTX 4060 Laptop GPU.

In the comprehensive evaluation of performance metrics, this study focuses on the structural failure risk assessment and adopts a dual-indicator co-optimisation strategy. The global prediction accuracy is quantified by the Mean Squared Error (MSE), complemented by the coefficient of determination ($R^2$) to evaluate the model's ability to capture the mechanical response. Priority is given to ensuring the MSE control of the prediction of the high-strain regions sensitive to structural failure, while the credibility of the full-field mechanical response law is guaranteed by the $R^2$ metric. This approach enables robust prediction of the strain field distribution characteristics of the grain, balancing computational efficiency and engineering accuracy. Unless otherwise specified, MSE and $R^2$ are the average values of four load cases.

### 4.2 Comparison of the optimisation effects of the Adaptive Top-k Pooling method

To verify the effectiveness of ATP, a series of comparative experiments is designed in this study to compare the performance of average pooling, ASAP pooling, SAG pooling, and adaptive Top-k pooling. These experiments are named SM-A, SM-B, SM-C, and SM-D, respectively.

As presented in Table 3 and Fig. 6, the experimental results on the strain field prediction dataset for the rear wing star hole grains demonstrate that the ATP method outperforms other comparative methods across all key metrics, exhibiting a significant overall advantage. In terms of prediction accuracy, the $R^2$ value for SM-D reaches 0.9284, approximately 0.9% higher than that of the second-best method, SM-C, and substantially higher than those of SM-A and SM-B. In terms of MSE metrics, SM-D achieves the lowest value at $2.0766 \times 10^{-4}$, approximately 12.3% lower than that of SM-C, further confirming its superior error control.

In terms of computational efficiency, the ATP method shows marked improvement over other methods. Its single-epoch training time is only 7.88 seconds, more than 10 times faster than the slowest method. Additionally, its temporal stability is higher, with a standard deviation of just 0.04, indicating excellent controllability and robustness, and achieving optimal overall performance.

Table 3 Performance comparison of pooling methods

| Pooling Method | $R^2$ | MSE ($\times 10^{-4}$) | Time/Epoch (s) |
|---|---|---|---|
| SM-A(Average) | 0.7234 | 5.2714 | 71.17 ± 17.37 |
| SM-B(ASAP) | 0.7666 | 4.9025 | 91.04 ± 0.84 |
| SM-C(SAG) | 0.9194 | 2.3689 | 88.90 ± 19.28 |
| **SM-D(ATP)** | **0.9284** | **2.0766** | **7.88 ± 0.04** |

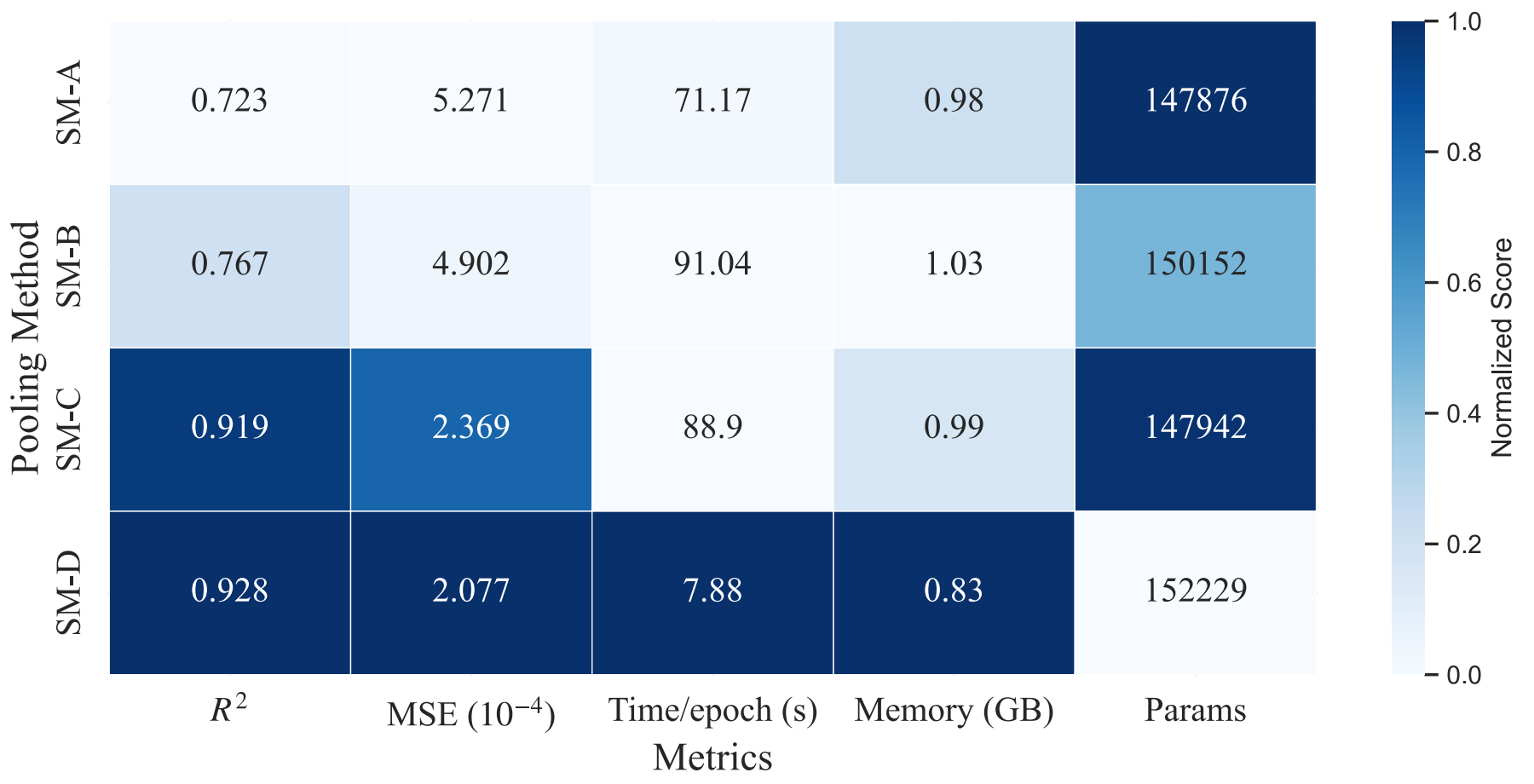

Fig. 6 Comparative Heatmap of Pooling Methods

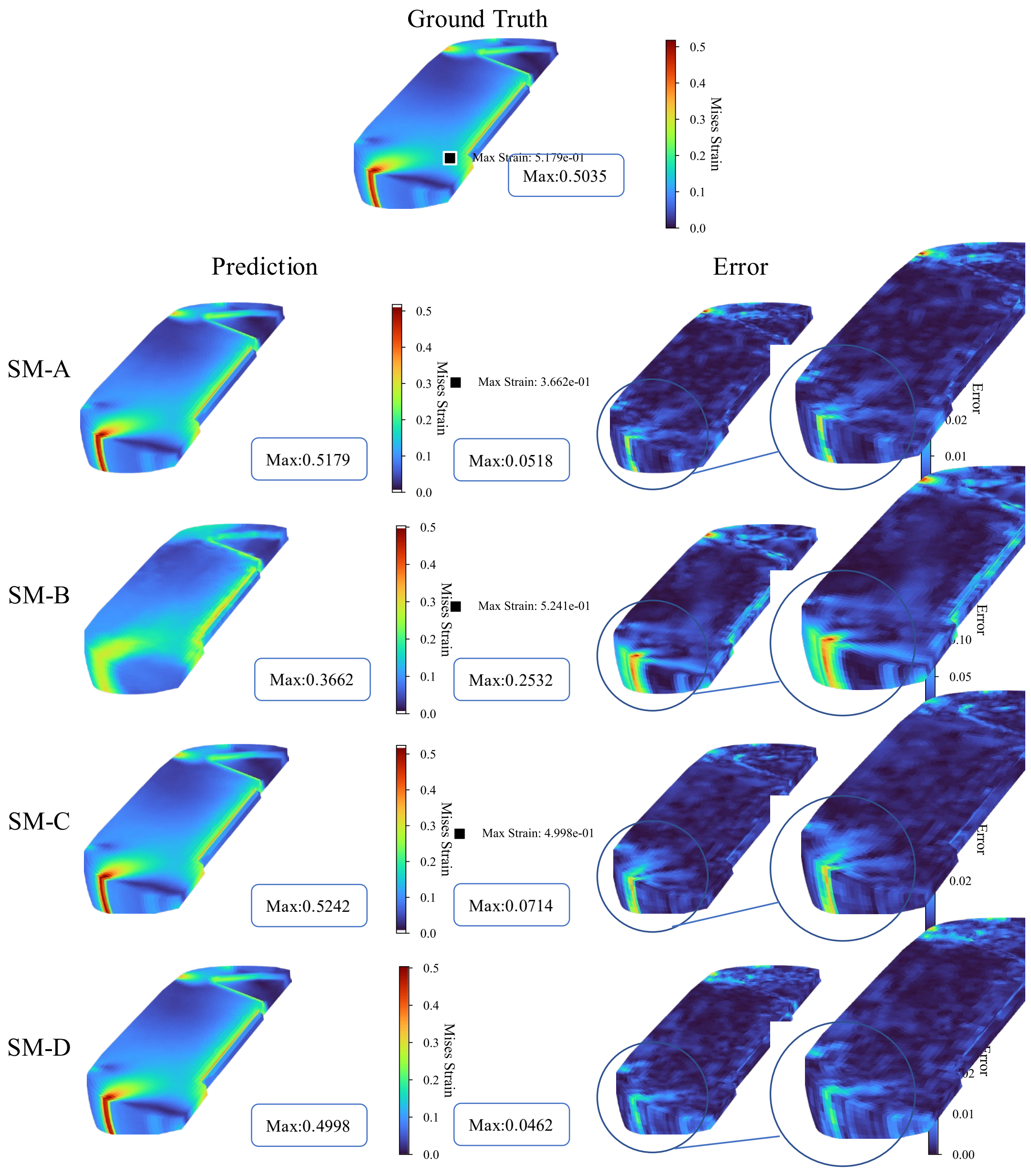

Fig. 7 Comparison of different pooling methods.

Fig. 7 illustrates the strain field prediction results for the grain under acceleration load condition, revealing that the maximum strain is concentrated in the anterior debonding seam region. All pooling methods, except SM-B, successfully identify the anterior and posterior debonding seams as critical regions. However, the SM-D method maintains a consistently low error level across the entire domain, particularly in the debonding seam regions, with a prediction error in the anterior debonding seam region 33.0% lower than that of the next-best method, SM-C.

From a mechanistic perspective, this performance stems from the unique mechanism of the ATP method, which dynamically learns the mechanical significance of nodes through a topology-independent MLP scoring network. This approach overcomes the feature smoothing of average pooling(SM-A), the local feature dilution of ASAP pooling（SM-B）, and the neighbourhood dependency and high-risk region pruning limitations of SAG pooling(SM-C). Consequently, ATP's dynamic feature selection mechanism effectively preserves critical features in high-risk regions, providing more reliable predictions for structural safety assessments.

### 4.3 Effect of the number of the Adaptive Feature Fusion layers on the results

To systematically evaluate the impact of feature fusion depth on prediction performance, all other hyperparameters are held constant, and the fusion depth is varied across 0–4 layers. Depths of 0 and 1 layer are considered non-fusion states and are collectively designated as the baseline model, FM-A, while the remaining depths are denoted as FM-B, FM-C, and FM-D, corresponding to increasing fusion depths.

Table 4 Performance comparison of fusion depths

| Model | $R^2$ | MSE ($\times 10^{-4}$) | Time (s)/Epoch | Parameter Count |
|---|---|---|---|---|
| FM-A | 0.8643 | 5.9785 | 7.45 ± 0.03 | 149,093 |
| FM-B | 0.9198 | 2.3546 | 7.71 ± 0.05 | 152229 |
| **FM-C** | **0.9284** | **2.0766** | **7.88 ± 0.04** | **152229** |
| FM-D | 0.9278 | 2.1751 | 8.53 ± 0.32 | 152229 |

The experimental results show a significant modulation effect of the AFF module on prediction accuracy. As shown in Table 4, FM-C (with three-layer fusion) achieves the optimal performance on the test set, with its MSE reduced to $2.0766 \times 10^{-4}$, and its $R^2$ improved to 0.9284, achieving a 65.3% reduction in MSE compared to the baseline model. This indicates that moderate cross-scale feature fusion can effectively enhance the model representation. However, when the fusion depth is increased to 4 layers (FM-D), the prediction accuracy declines. This phenomenon is directly attributed to the fusion mechanism that reuses a fixed MLP, leading to exponential decay of high-frequency information. Additionally, irrelevant noise from lower layers is homogenously propagated to higher layers, causing effective features to be overwhelmed.

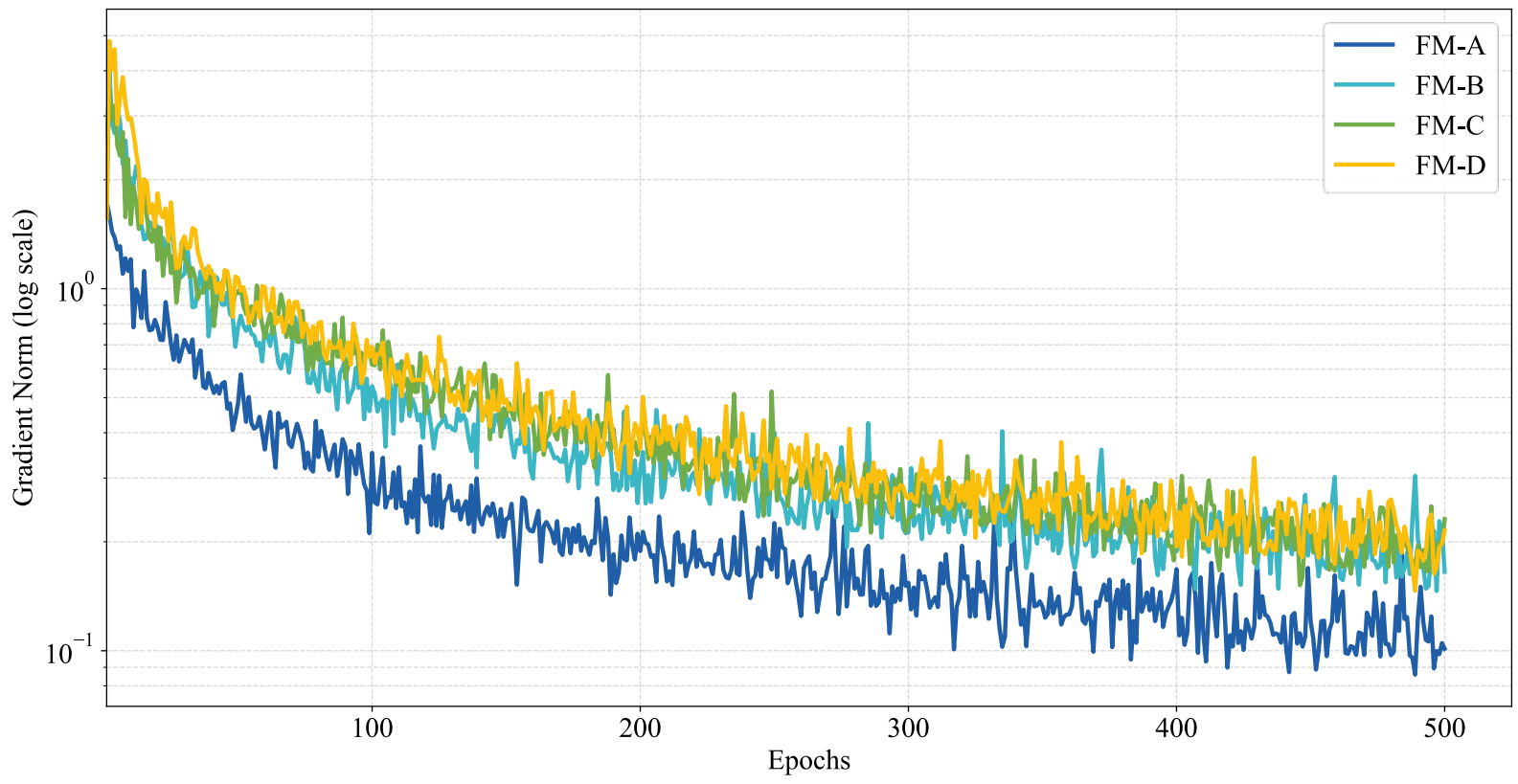

Fig. 8 Gradient Decay Process.

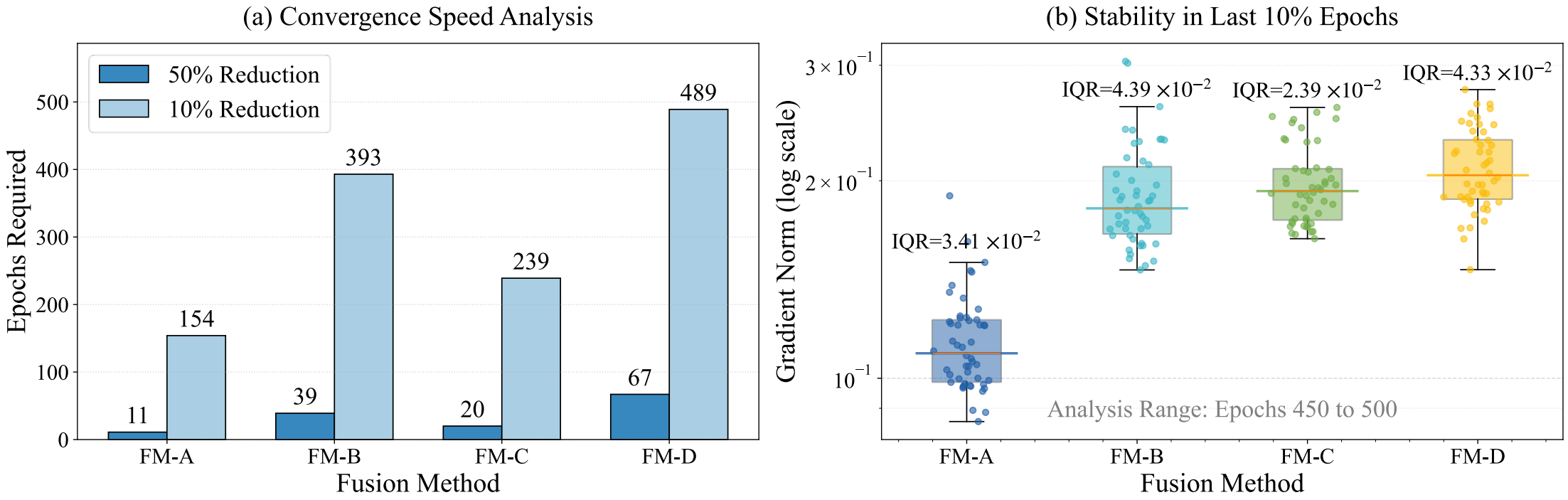

Fig. 9 Gradient Norm Comprehensive Analysis.

To further investigate the intrinsic mechanisms during training, Fig. 8 illustrates the dynamic evolution of the gradient L2 norm during the training process. The analysis shows that the feature fusion mechanism effectively promotes the backpropagation of the gradient and alleviates the gradient decay problem by establishing a cross-layer information pathway. In terms of convergence characteristics, FM-C shows dual advantages. Firstly, the convergence speed is significantly improved: 50% gradient attenuation is achieved at the early stage of training (within 20 epochs), and its convergence time is shortened by 70.1% compared with that of FM-D (Fig. 9(a)). Secondly, the training stability is enhanced: at the late stage of training, the interquartile range (IQR) of gradient fluctuation is reduced to $2.39 \times 10^{-2}$ , which is 29.9% lower than that of FM-A (Fig. 9(b)). This indicates that the three-layer fusion structure achieves an optimal balance between convergence efficiency and training robustness.

In summary, the feature fusion mechanism markedly enhances the model's learning capacity by integrating information across multi-scale feature layers; however, its effectiveness is highly contingent on the choice of fusion depth. The present experiments identify three layers as the optimal fusion depth.

### 4.4 Ablation experiment

To evaluate the independent contributions of the ATP and AFF modules, this study designs a systematic ablation experiment. Four configurations are compared using the control variable method: the baseline model (Model-A), a traditional graph U-Net model; Model-B, retaining only the AFF module; Model-C, retaining only the ATP module; and Model-D, the complete model.

In terms of module-independent contribution, the ATP module (Model-B) resulted in a significant improvement in prediction accuracy, with a 10.3% increase in $R^2$ and a 19.2% reduction in MSE, with a mechanism that optimises the retention of nodes in key regions. The AFF module (Model-C) contributes greater performance gains (19.5% improvement in $R^2$ and 40.7% reduction in MSE), validating the effectiveness of cross-scale feature integration for modelling complex mechanical fields.

From the synergy of the complete model (Model-D), $R^2$ is improved by 28.3% and MSE is reduced by 62.8% compared to Model-A, and outperforms the sum of the independent effects of the modules (Table5, Fig. 10). Notably, the full model is trained approximately 7 times faster than the baseline model while maintaining the highest prediction accuracy, despite the 5.2% increase in the number of parameters. This is attributed to the complementary mechanism of the dual modules: the hierarchical structure constructed by adaptive sampling optimises the feature fusion path, while the fused feature representation feeds the sampling decision.

Table 5 Comparison of ablation experiment results

| Model | $R^2$ | MSE ($\times 10^{-4}$) | Time/Epoch (s) | Parameter Count |
|---|---|---|---|---|
| Model-A | 0.7234 | 5.5879 | 56.14 ± 0.13 | 144,740 |
| Model-B | 0.7976 | 4.5154 | 7.99 ± 0.66 | 154,374 |
| Model-C | 0.8643 | 3.3142 | 24.49 ± 25.94 | 149,093 |
| **Model-D** | **0.9284** | **2.0766** | **7.88 ± 0.04** | **152229** |

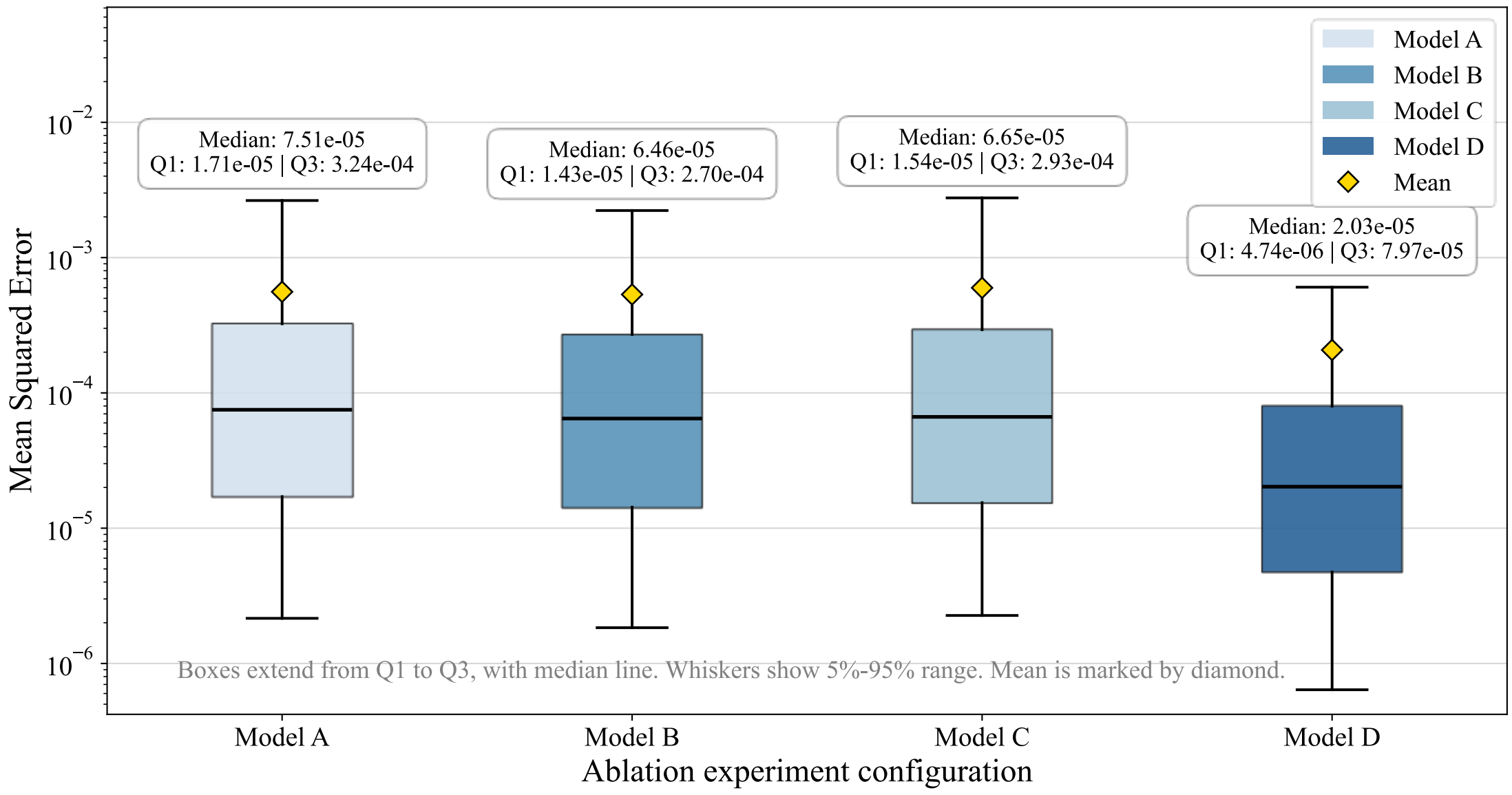

Fig. 10 Comparison of ablation experiment results

## 4.5 Comparison of prediction accuracy under multiple load conditions

The study predicts the steady-state strain field of the propellant structure under four operating conditions: curing and cooling (Load Case 1), vertical storage (Load Case 2), ignition pressurisation (Load Case 3), and overload (Load Case 4). As shown in Fig. 11, the prediction accuracy varies significantly across conditions: Load Cases 1 and 2 achieve $R^2$ of 0.9557 and 0.9548, respectively, markedly outperforming Load Case 3 (0.8982) and Load Case 4 (0.9069). The MSE and PAE errors for Load Cases 1 and 2 are an order of magnitude lower than those for Load Cases 3 and 4, as further validated by the three-dimensional strain field distributions in Fig. 12-13. Notably, the multi-view error distribution in Fig. 13 reveals that the maximum MSE concentrates at the front and rear debonding seams. Fig. 14(e) illustrates the extraction of points along the outer boundary, forming a characteristic line by rotating counterclockwise from the starting point. Fig. 14 (a)–(d) compare the predicted and ground truth values along this line under different conditions, indicating that transient conditions (Load Cases 3 and 4) exhibit more complex strain distribution patterns and higher peak strains at structural boundaries.

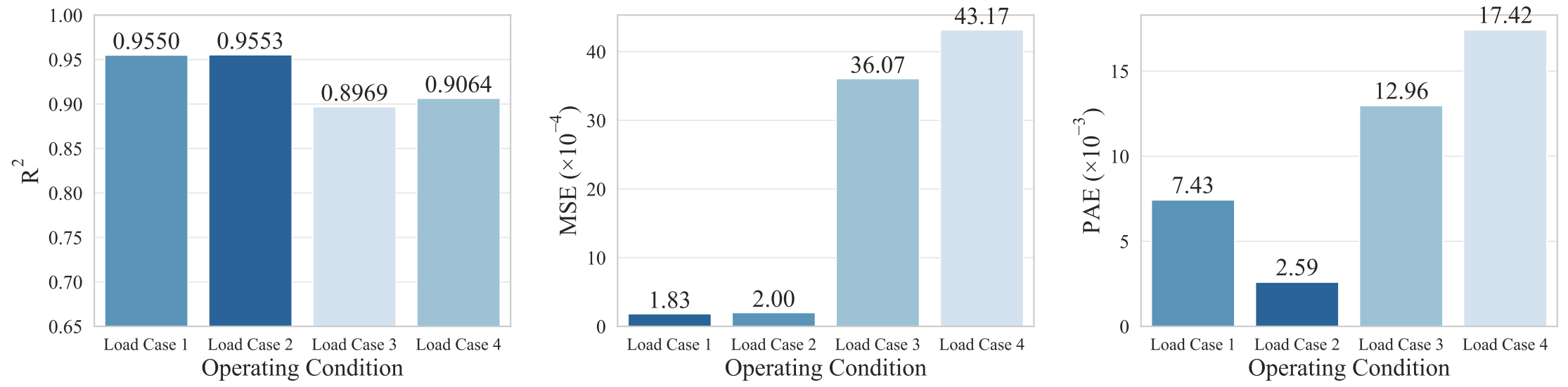

Fig. 11 Comparison of prediction accuracy under different load cases

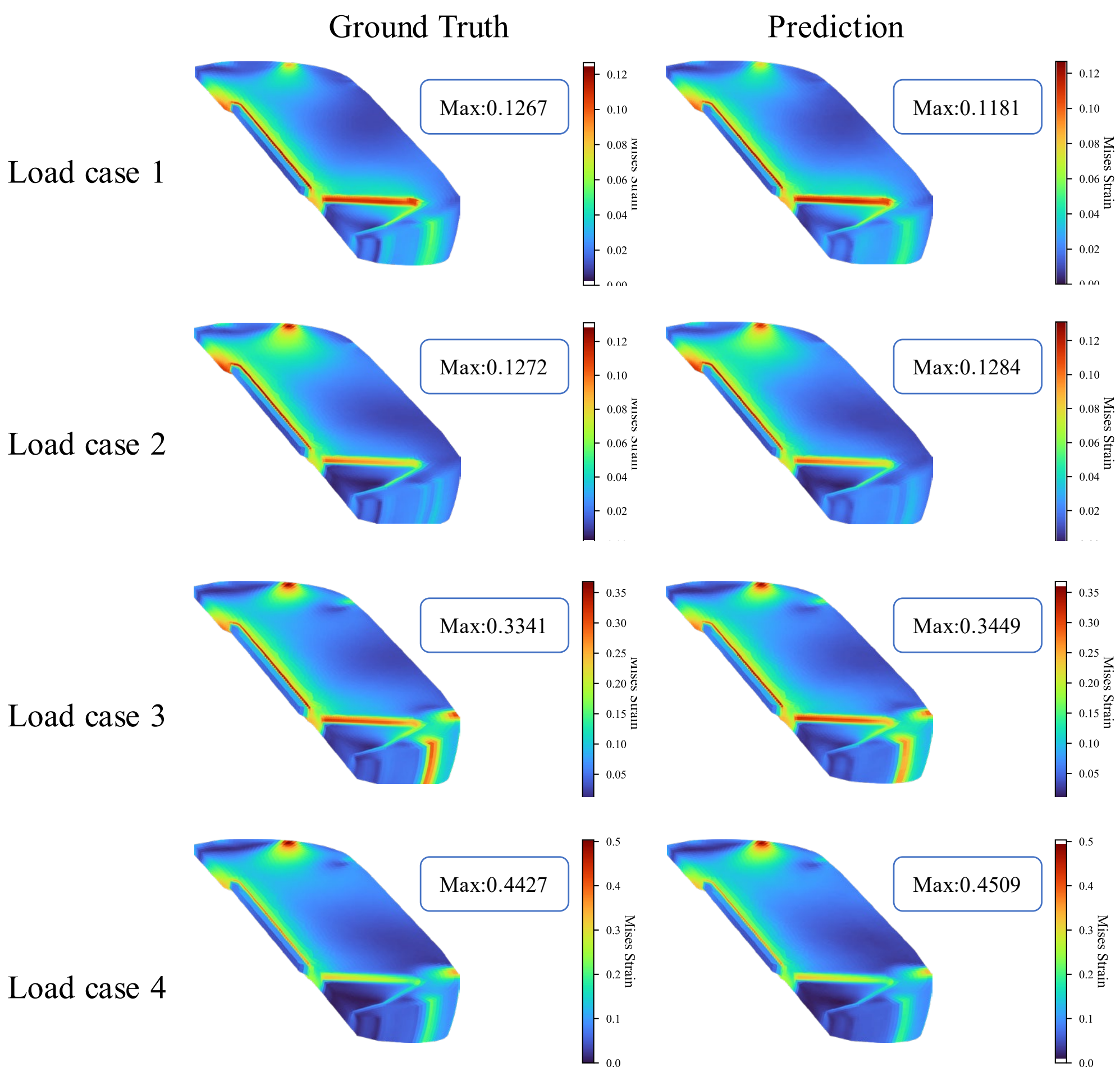

Fig. 12 Comparison under different load cases.

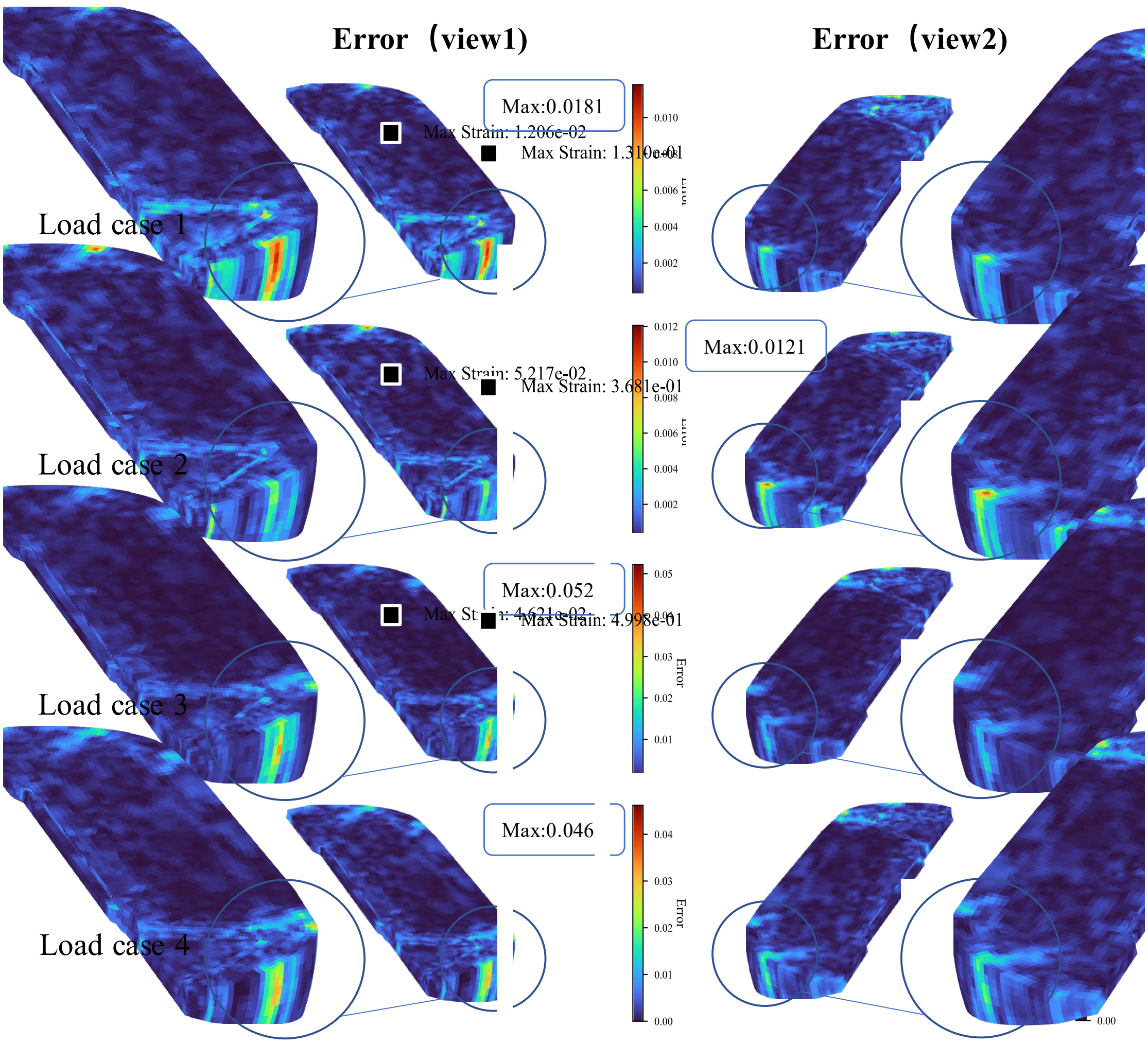

Fig. 13 Comparison of errors under different load cases.

From a physical perspective, the differences in prediction accuracy stem from the adaptability of the network architecture to load characteristics. Although the ATP module effectively captures the global strain distribution under quasi-static conditions, its reliance on absolute strain values for importance assessment leads to systematic undersampling of high-gradient, low-amplitude regions, such as debonding seams, resulting in the loss of strain gradient information under transient conditions. Concurrently, the AFF module introduces mapping distortions in strain-discontinuity regions due to upsampling of clustering indices during cross-scale transmission, while the fixed-weight fusion MLP struggles to adapt to the spectral properties of transient loads. These local errors propagate through the decoder's residual paths, leading to significant error accumulation at structural discontinuities.

Building on insights into the underlying mechanisms, future research could explore: (1) adaptive sampling methods using strain gradient tensors to improve feature capture in high-strain-rate regions; (2) dynamic fusion modules to enhance cross-scale feature transfer; and (3) physical equation constraints to mitigate cascading errors in critical regions. These advancements aim to enhance modelling accuracy under complex loading conditions and advance solid rocket motor structural analysis.

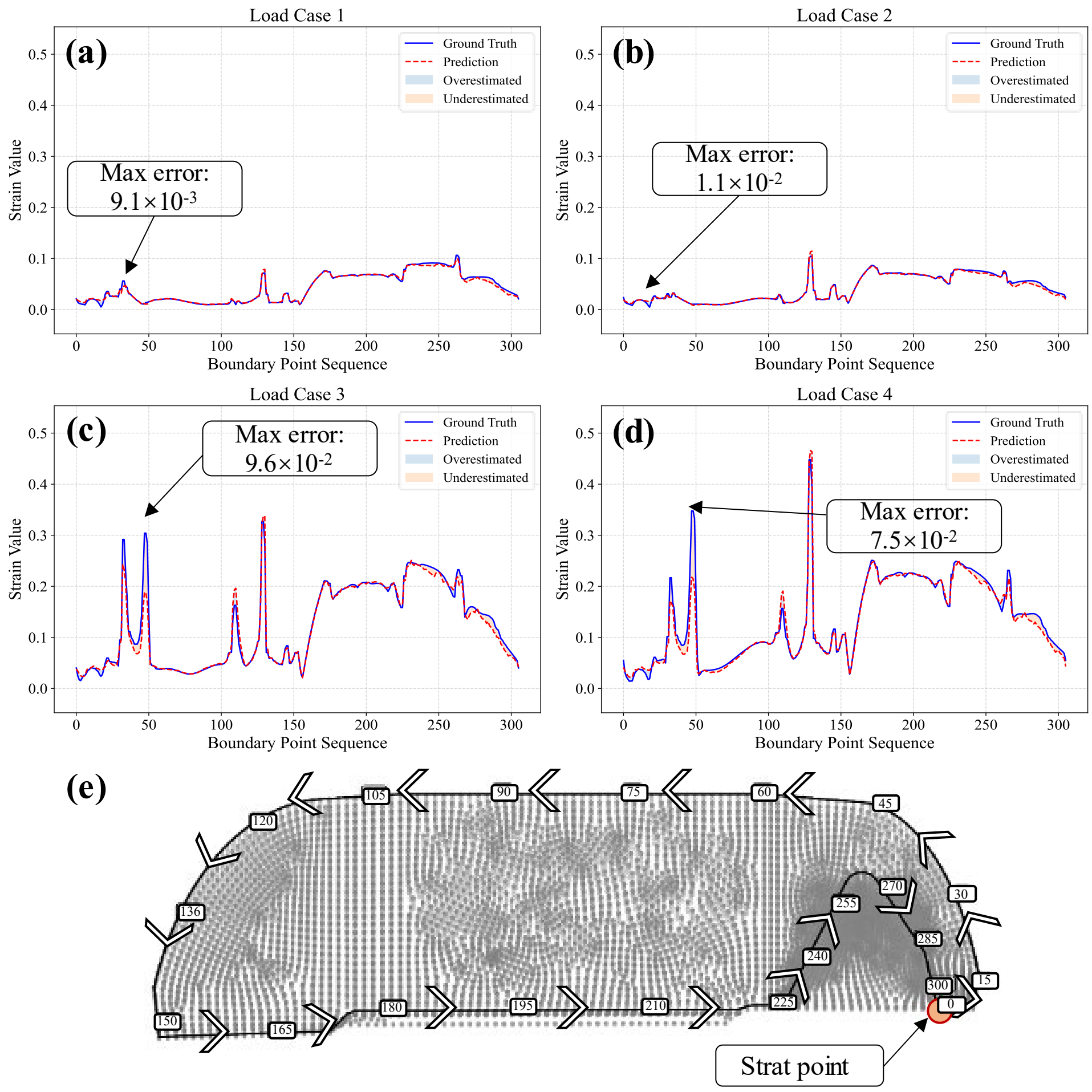

Fig. 14 Comparison of predicted and ground truth values on the characteristic line under different load cases

## 5. Conclusions

This paper proposes GrainGNet, an adaptive graph neural network framework designed for rapid, high-precision strain field prediction in solid rocket motor grains. By integrating the Adaptive Top-k Pooling (ATP) module with the Adaptive Feature Fusion (AFF) module, it achieves efficient full-field strain modelling for complex geometric grain structures under diverse loading conditions. Experimental results demonstrate:

1) The ATP method employs a lightweight MLP for nonlinear node importance assessment, reducing prediction errors in critical regions, such as debonding seams, by 33% compared to the next-best pooling method. It also reduces single-epoch training time to 7.88 seconds, a tenfold efficiency improvement, enabling efficient structural safety assessments. The AFF module, using a three-layer pyramid structure, optimises cross-scale information transfer, achieving a 65.3% reduction in MSE compared to the baseline model, indicating that moderate cross-scale feature fusion significantly enhances representational capacity.

2) Ablation studies confirm the synergistic effect between the two modules, with the full model reducing total error by 62.8% compared to the traditional U-Net, surpassing the combined independent effects of the modules. With only a 5.2% increase in parameters, the training efficiency exceeds the baseline model by sevenfold, reflecting a beneficial interplay between hierarchical structure and feature enhancement.

3) Analysis of operating condition adaptability shows that the model achieves significantly higher prediction accuracy under quasi-static conditions than under transient conditions. This indicates that the current sampling strategy is insufficiently sensitive to high-gradient, low-strain regions, requiring physical constraints to enhance applicability under complex loading conditions.

**Declaration of competing interest**

There is no conflict of interest to declare in the article.

**Acknowledgments**

This work was supported by the National Natural Science Foundation of China (12372203, 12302028)